\documentclass[aps,prl,reprint,superscriptaddress,letterpaper,twocolumn,longbibliography,floatfix]{revtex4-2}
\usepackage[english]{babel}
\usepackage[T1]{fontenc}

\usepackage{graphicx}
\usepackage{dcolumn}
\usepackage{siunitx}
\usepackage{amsmath}
\usepackage{amssymb}
\usepackage{amsthm}
\usepackage{amsfonts}
\usepackage{mathtools}
\usepackage{mathrsfs}
\usepackage{graphicx}
\usepackage{bm}
\usepackage{bbm}
\usepackage{empheq}
\usepackage{cases}
\usepackage{euscript}
\usepackage[svgnames,dvipsnames,x11names,table]{xcolor}
\usepackage[colorlinks=true,linkcolor=SpringGreen4,citecolor=Mulberry,urlcolor=Magenta3]{hyperref}
\usepackage[shortlabels]{enumitem}
\usepackage[draft,todonotes={textsize=scriptsize}]{changes}
\setuptodonotes{inline}

\usepackage{booktabs}
\usepackage{upgreek}
\definechangesauthor[name={Eva Weig}, color=violet]{EW}
\definechangesauthor[name={Hugo Ribeiro}, color=SpringGreen4]{HR}
\definechangesauthor[name={Avishek Chowdhury}, color=Blue4]{AC}
\definechangesauthor[name={Anh-Tuan Le}, color=RedOrange]{ATL}
\newcommand{\bra}[1]{\langle #1|}
\newcommand{\ket}[1]{|#1\rangle}

\newcommand{\ketbra}[2]{\ket{#1}\!\bra{#2}}

\newcommand{\mm}[1]{\mathrm{#1}}

\newcommand{\lr}[1]{\left(#1\right)}
\def \ud{\mathrm{d}}

\def \uf{\mathrm{f}}

\def \ui{\mathrm{i}}

\def \us{\mathrm{s}}

\def \uw{\mathrm{w}}

\def \np{\rotatebox[origin=c]{45}{$\Leftrightarrow$}}
\def \nm{\rotatebox[origin=c]{-45}{$\Leftrightarrow$}}
\def \Ev{\mbox{\boldmath$E$}}

\def \Ev{\mbox{\boldmath$E$}}

\DeclareFontFamily{OT1}{pzc}{}
\DeclareFontShape{OT1}{pzc}{m}{it}{<-> s * [1.10] pzcmi7t}{}
\DeclareMathAlphabet{\mathpzc}{OT1}{pzc}{m}{it}
\DeclareSIUnit\bar{bar}
\DeclareSIUnit{\belmilliwatt}{Bm}
\DeclareSIUnit{\dBm}{\deci\belmilliwatt}

\begin{document}

\title{Precise estimation of the coupling strength between two nanomechanical modes from four Ramsey fringes}

\author{Anh Tuan Le}
\affiliation{School of Computation, Information and Technology, Technical University of Munich, 85748 Garching, Germany}
\author{Avishek Chowdhury}
\affiliation{School of Computation, Information and Technology, Technical University of Munich, 85748 Garching, Germany}
\author{Hugo Ribeiro}
\affiliation{Department of Physics and Applied Physics, University of Massachusetts Lowell, Lowell, MA 01854, USA}
\author{Eva M. Weig}
\affiliation{School of Computation, Information and Technology, Technical University of Munich, 85748 Garching, Germany}
\affiliation{TUM Center for Quantum Engineering (ZQE), 85748 Garching, Germany}
\affiliation{Munich Center for Quantum Science and Technology (MCQST), 80799 Munich, Germany}

\begin{abstract}
We experimentally determine the coupling strength between two strongly coupled nanomechanical modes using a Ramsey-inspired
technique optimized for signals as short as four fringes. The method is applied to precisely probe the change of the coupling rate
induced by a modification of the microwave-cavity readout field. It opens a pathway towards sensing electrostatic field
fluctuations approaching single-charge resolution.
\end{abstract}

\maketitle

\textit{Introduction ---}
Fluctuations in physical and biological systems are an ubiquitous phenomenon that strongly impacts their
behavior~\cite{Giessibl2003,Paladino2014,Tsimring2014,Degen2017,Pezze2018,Bachtold2022}.  They govern physical processes,
including decoherence~\cite{Clerk2010}, and limit the sensitivity of detectors~\cite{Bachtold2022}. There is a range of
established techniques for their characterization~\cite{Maizelis2011,Sun2015,Sansa2016,Wang2020}, which all rely on probing
spectral properties.  This is sufficient to probe Gaussian noise, the statistics of which is fully determined by its average and
its spectrum, i.e. the first and second moment.  However, a full characterization of the noise in a system requires, in general,
knowledge of all its moments~\cite{Wudarski2023}. This is particularly important for non-Gaussian noise, where higher-order
moments are required to obtain the full statistics. Probing higher-order moments remains an experimentally outstanding
challenge~\cite{Gershon2008, Christensen2019,Sung2019,White2025,Curtis2025}.

Another, elegant, approach would be to time-resolve the fluctuations directly. This calls for fast and sensitive probes.
Time-resolved measurements of fluctuating electromagnetic quantities have been
demonstrated~\cite{Schoelkopf1998,Gustavsson2008,Klimov2018,Bartolomei2025}. These techniques are, however, limited to cryogenic
temperatures.  Here we present a scheme that allows to time-resolve fluctuations at room temperature. It exploits Ramsey
spectroscopy to probe the frequency splitting between the hybridized modes in a nanomechanical two-mode system~\cite{Faust2013}.
Using the recently proposed iterative adaptive spectroscopy protocol~\cite{Chowdhury2022}, the splitting magnitude is accurately
obtained from only a small number of Ramsey fringes. We confirm that the splitting depends on the electrostatic
environment~\cite{Rossi2016a, Lepinay2017,Braakman2019}. This opens the way towards investigating non-Gaussian charge fluctuations
at room temperature~\cite{Wudarski2023}, with an estimated single-electron sensitivity and sub-millisecond time resolution.

\begin{figure}[t]
    \includegraphics[width=\columnwidth]{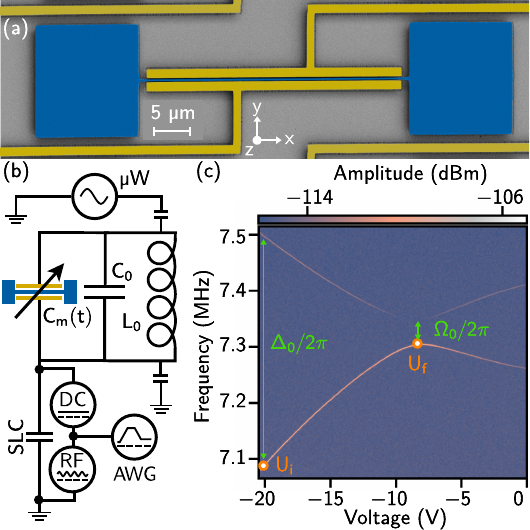}
    \caption{\textbf{Experimental realization}. \textbf{(a)} Scanning electron micrograph of a typical silicon nitride string
        resonator (blue) with the two gold electrodes nearby. \textbf{(b)} Simplified measurement setup. See main text for
        details.  \textbf{(c)} Color plot of amplitude spectra measured as a function of DC voltage, showing an avoided crossing
        of magnitude $\Omega_0$ between the OOP and the IP mode. The two orange circles denote the DC voltage used for
        initialization and for tuning on resonance, respectively. $\Delta_0$ is the frequency difference of the two mechanical
        modes at $U_{\text{i}}$.}
    \label{fig:01}
\end{figure}

\textit{Nanomechanical two-mode system ---} 
The nanomechanical two-mode system is implemented using the two orthogonally polarized fundamental flexural eigenmodes of a
nanomechanical string resonator. They are associated with the motion of the resonator along the y and z-axis, vibrating at
$\omega_j/(2\pi) \approx 7\,\si{\mega\hertz}$ where $j\in \{\mm{IP}, \mm{OOP}\}$ for the out-of-plane (OOP) and in-plane (IP)
mode, respectively. A typical geometry  of the system is shown using a scanning electron micrograph in Fig.~\ref{fig:01}(a)
[see also Appendices~\ref{app:oop-ip} and \ref{app:2osc}]. The $55\,\mu$m-long nanostring resonator is made of
strongly pre-stressed silicon nitride, and exhibits quality factors in the range of $250,000$ at room temperature and
$10^{-4}\ \si{\milli\bar}$. The resonator is flanked by two gold electrodes used for dielectric control. A schematic of the
measurement setup is depicted in Fig.~\ref{fig:01}(b); a more detailed description is found in
Appendices~\ref{app:3Dcavity} and \ref{app:setup}. One
of the electrodes is connected to an antenna that couples to a coaxial $\lambda/4$ three-dimensional (3D) microwave cavity
(see Appendix~\ref{app:3Dcavity}) used for displacement detection \cite{Faust2012a, Reagor2013, Le2021}. The 3D cavity is made
of copper, suitable for room-temperature operation. It is driven by a microwave generator ($\mu$W) on resonance
$\omega_\mm{c}/( 2\pi)\approx 3\ \si{\giga\hertz}$ at $P_\mm{c}=22\ \si{\dBm}$.  In Fig.~\ref{fig:01}(b), the 3D microwave
cavity is represented by a tank circuit with inductor $L_0$ and capacitor $C_0$. The second electrode is wire-bonded to a a
single layer capacitor (SLC) that provides a microwave ground path~\cite{Rieger2012} while connecting to a self-made voltage
combiner adding a DC and a RF signal with the output of an arbitrary waveform generator (AWG).

The mechanical motion periodically modulates the capacitance $\mathrm{C_m} (t)$ of the electrodes, inducing sidebands
in the cavity response at frequencies $\mathrm{\omega_c\pm\omega_j}$ which are demodulated using heterodyne IQ-mixing as
described in Appendix~\ref{app:setup}.  Application of a DC voltage to the second electrode allows for dielectric tuning of the
mechanical eigenfrequencies $\omega_j$ \cite{Unterreithmeier2009, Rieger2012}. The DC voltage also induces strong dielectric
coupling between the IP and OOP mode the magnitude of which is governed by the polarizability of the dielectic material $\alpha$
and the inhomogeneous electric field $\Ev$ between the electrodes~\cite{Faust2012a} (see also Appendix~\ref{app:2osc}), This gives
rise to a pronounced avoided level crossing [see Fig.~\ref{fig:01} (c)] with a frequency splitting $\Omega_0$ around which both
modes hybridize into normal modes. The system can be described as a two-mode system, as all higher-order flexural modes only
appear at eigenfrequencies approximately given by $N\, \omega_j$, corresponding to a quasi-infinite anharmonicity since
$\Omega_0/(N\, \omega_j) \ll 1$  for $N\geq 1$.  Combination of the DC voltage with a RF voltage resonant at $\omega_j$
dielectrically actuates mode $j$ \cite{Unterreithmeier2009}.   

\textit{Ramsey-based Iterative Adaptive Sensing (IAS) ---} 
The straightforward way to measure the normal mode splitting in our system is to perform a spectroscopic noise driven measurement
of the resonant frequencies as a function of applied DC voltage [see Fig.~\ref{fig:01} (c)] and extract the value of $\Omega_0$
from a fit. The fit function is chosen to match the spectrum of two classically coupled harmonic oscillators \cite{Novotny2010}
(see also Appendix~\ref{app:2osc}), which provides a good description of the spectrum in the vicinity of the
avoided crossing. However, the robustness of the fit depends heavily on the signal to noise ratio of the measured data and the
precise knowledge of several system parameters. Additionally, such spectroscopic measurements are based on parametric sweeps and
require long average times, making them unsuitable for sensing applications.

The alternative relies on the observation that at the avoided crossing, the hybridized modes can exchange energy at a rate faster
than the characteristic decoherence time $T_\ud$ of the system. Thus, in the limit $\Omega_0/(2\pi) T_\ud \ll 1$, the hybridized
modes can be coherently controlled, allowing for the implementation of Ramsey spectroscopy-based sensing
\cite{Faust2013,Gittsovich2015,Cleland2024,Yang2024}. 

Applied to our system, it consists of the following steps also visualized in Fig.~\ref{fig:02} (see also
Appendix~\ref{app:ramsey}):
\begin{figure}[t]
    \includegraphics[width=\columnwidth]{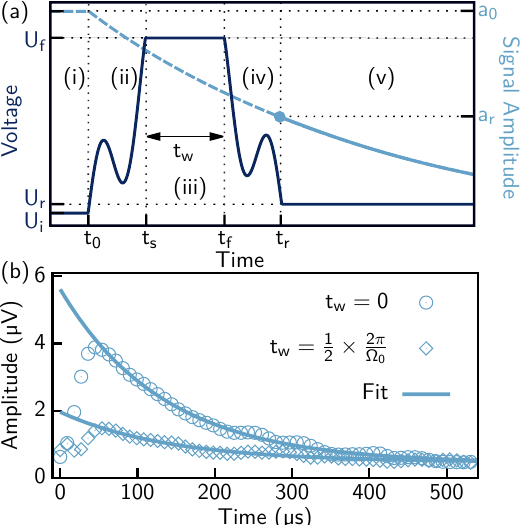}
    \caption{\textbf{Implementation of the Ramsey interferometry}. \textbf{(a)} Ramsey pulse sequence with the corresponding first
        harmonic Magnus-based correction of the leading (ii) and the trailing (iv) edges for frequency-sweep. Note we introduce a
        finite voltage offset $U_r-U_i\neq 0$ that allows the readout state $\boldsymbol{a}_r$ to decay, whereas the initial state
        $\boldsymbol{a}_i$ is continuously driven. \textbf{(b)} Ringdown measurement of the IP mode after completing the Ramsey
        sequence for two different waiting times $t_\uw$. The exponential fits extrapolate back to $t_0=0$ to compensate the
        mechanical damping during the evolution time and are used to convert the signal into a return probability~\cite{Seitner2016,
        Seitner2017}.}
    \label{fig:02} 
\end{figure}
\begin{enumerate}[(i)]
    \item Initialize the IP mode by applying a RF voltage at $\omega_\mm{IP}(U_\ui)$ at a DC bias $U_\ui$ [see Fig.~\ref{fig:01}
        (c)] fulfilling the condition $\Delta_0 = \omega_\mm{OOP} - \omega_\mm{IP} \gg
        \Omega_0$.
    \item Using the initial estimate $\bar{\Omega}_{0}$ of $\Omega_0$ obtained from the spectrum fit, design a non-adiabatic
        frequency-sweep $\Delta (t)$ that quickly brings the IP mode into the avoided level crossing at a DC bias of $U_\uf$ with
        a high-fidelity [see Fig.~\ref{fig:02} (a)]. A voltage signal encoding $\Delta (t)$ is generated by an Arbitrary Waveform
        Generator (AWG) and applied to the system along with the DC and RF voltage [see Fig.~\ref{fig:01} (b)]. This prepares the
        ideal sensing state \cite{Chowdhury2022}. In general, the ideal sensing state consists of an equal superposition of the
        two normal modes. At the avoided crossing, the normal modes vibrate with a polarization of $\pm
        45\mbox{\textdegree}$~\cite{Faust2012a,Faust2013}. Here we choose the equal superposition reproducing the IP mode
        [see Appendix~\ref{app:ramsey}].  
  \item Let the system evolve freely for a time $t_\uw=2\pi n/\bar\Omega_0$, where $n$ is an integer. The coherent evolution
      allows this state to pick up a phase equal to $\phi=\bar\Omega_0 t_\uw$.
  \item Bring the system back to a bias $U_\mm{r}$ close to the initial bias point $U_\ui$ by applying the reverse frequency-sweep
      to $\Delta (t)$ defined in (ii).  Ideally, $U_\mm{r}$ should coincide with $U_\ui$. In practice, a finite offset of a few
      $100$\,mV is applied to avoid interference with the RF drive voltage [see Fig.~\ref{fig:02} (a)].
  \item Extract the amplitude of the IP mode at $U_\mm{r}$ from a ringdown measurement as a function of the free evolution time
      $[0,t_\uw]$ [see Fig.~\ref{fig:02} (b)] and convert it into a return probability~\cite{Seitner2016, Seitner2017}. This
      constitutes the Ramsey signal. 
\end{enumerate}

The obtained amplitude as a function of time yields an ideal Ramsey signal with unity visiblity ignoring decoherence and assuming
perfect knowledge of the splitting $\Omega_0$. However the spectroscopic method does not provide that level of accuracy, since it
involves a small but non-negligible error in the determination of $\Omega_0$, which severely deteriorates the Ramsey signal as
described in Ref.~\cite{Chowdhury2022}. The recently proposed IAS (Iterative Adaptive Spectroscopy) protocol~\cite{Chowdhury2022}
provides a means to bypass both decoherence and the imprecise knowledge of $\Omega_0$.  At its heart IAS represents an enhanced
Ramsey protocol that allows one to precisely extract frequency information from short Ramsey signals consisting only of a small
number of fringes $n$ ($n<10$). It combines an iterative procedure with signal processing to achieve a fast and accurate
measurement of the frequency splitting of a two-mode system. Using the spectroscopic guess as a prior estimate of the splitting
$\bar{\Omega}_{0}^{(0)}$, we perform a first Ramsey sequence [(i) - (v)]. A fast Fourier transform of the short Ramsey signal is
performed to extract a new estimate $\bar{\Omega}_{0}^{(1)}$. This constitutes the starting point of IAS.  Using
$\bar{\Omega}_{0}^{(1)}$, all input parameters of the Ramsey sequence are updated before proceeding to the next iteration $m=2$
(see Appendix~\ref{app:pulse}). To avoid known issues linked to Fourier transforms of short signals, IAS relies on
signal processing (windowing and zero-padding) as described in \cite{Chowdhury2022}. This procedure is repeated until the
frequency estimate $\bar{\Omega}_{0}^{(m)}$ saturates.

Here, we experimentally apply the IAS protocol to the classical nanomechanical two-mode system
\cite{Novotny2010,Faust2013,Frimmer2014,Seitner2016,Seitner2017} to sense charge fluctuations
building up on the surface of the nanomechanical resonator. Notice that all data shown in
the following has been obtained after averaging over $30$ measurements.

\begin{figure}[t]
    \includegraphics[width=\columnwidth]{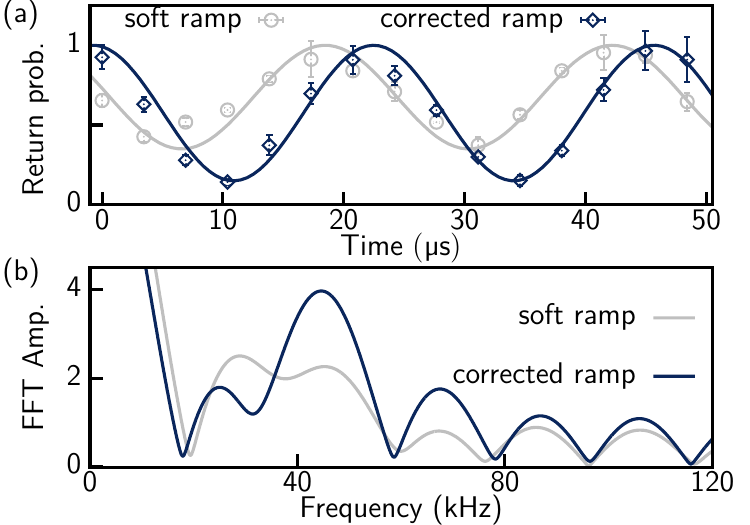}
    \caption{\textbf{State preparation}. \textbf{(a)} Ramsey signal (return probability of the state $\boldsymbol{a}_i$) for two
        different initialization protocols. Measurement time captures approximately $2$ periods. Data has been averaged over $30$
        measurements. Grey (blue) datapoints bar are obtained with a sinusoidal soft ramp (corrected ramp, c.f.~\ref{fig:02}).
        Solid grey (blue) lines indicate fits. \textbf{(b)} Fast Fourier transform of the short, zero-padded Ramsey signals. Grey
        (blue) line corresponds to the initialization via the soft (corrected) ramp, respectively.}
    \label{fig:03NEW} 
\end{figure}

\begin{figure}[t]
    \includegraphics[width=\columnwidth]{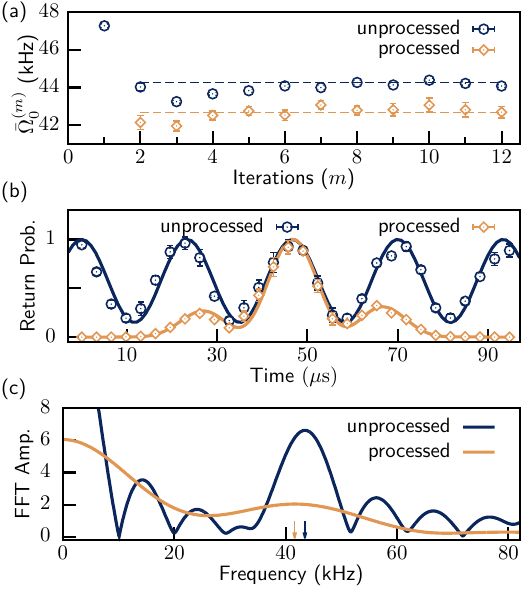}
    \caption{\textbf{Iterative adaptive spectroscopy for short and finite signals}. \textbf{(a)} Frequency estimation for each
        iteration step $m$. The $m=1$ datapoint was obtained from a two-period Ramsey signal as in Fig.~\ref{fig:03NEW}. Blue (orange)
        symbols refer to the unprocessed (processed) data. Error bars indicate standard deviation of the $30$ measurements. \textbf{(b)}
        Ramsey signal (return probability of the state $\boldsymbol{a}_i$) sampling $n=4$ fringes at $m=8$ before (blue) and after data
        processing (orange). \textbf{(c)} Fast Fourier transform of the zero-padded data shown in (b). Orange (blue) arrow marks position
        of the local maximum indicating frequency estimate $\bar\Omega_0^{(8)}$ in (a).}
    \label{fig:03}
\end{figure}

\textit{Precision frequency sensing with IAS ---} 
To illustrate the shortcomings of a standard Ramsey sequence, we measure the return probability of a non-iterated sequence over
$n=2$ fringes. The black data in Fig.~\ref{fig:03NEW} (a) and (b) depicts the Ramsey time trace as well as its fast Fourier
transform (FFT) for the case of a fast frequency sweep consisting of a simple cosine-shaped soft ramp~\cite{Chowdhury2022}. The
two apparent Ramsey fringes exhibit a poor visibility of only approximately $60$\,\%, and the FFT does not reveal a clear maximum
indicative of the splitting $\Omega_{0}$. The apparent double peak is an artefact from the intense side lobes arising from
finite-time effects (spectral leakage)~\cite{Chowdhury2022}. Note that ``maximum'' refers to a local maximum, excluding the
zero-frequency peak. We will employ this convention throughout. The blue data in Fig.~\ref{fig:03NEW} (a) and (b) was measured for
an improved frequency sweep based on a first harmonic Magnus-based correction [see Fig.~\ref{fig:02} (a)]. The visibility of the
Ramsey fringes for the corrected ramp is significantly improved to about $85$\,\%, and the FFT yields a clear maximum at
$\bar{\Omega}_{0}^{(1)}$. This is close to optimal, since the calculation of the corrected pulse relies on the inaccurate prior
estimate $\bar{\Omega}_{0}^{(0)}=41.3$\,kHz (see Supplemental Material~\ref{app:2osc}).  In addition, the spectral leakage still
distorts the spectrum, preventing to reliably extract the maximum. Notice that both traces extend slightly beyond the second
Ramsey fringe. This is also a consequence of the discrepancy between $\bar{\Omega}_{0}^{(0)}$ and $\Omega_{0}$. 

The IAS protocol is applied in Fig.~\ref{fig:03}. It displays a full IAS sequence for up to $m=12$ iterations. The number of
measured Ramsey fringes for $m>1$ is increased to $n=4$ in order to apply the IAS signal processing
procedure~\cite{Chowdhury2022}.  The extracted splitting $\bar{\Omega}_{0}^{(m)}$ as a function of the iteration $m$ is shown in
Fig.~\ref{fig:03} (a). The first datapoint $\bar\Omega_0^{(1)}$ is obtained from a FFT of a two-fringe time trace obtained with
the corrected ramp like in Fig.~\ref{fig:03NEW} (b), albeit frequencies do not match as both traces have been
obtained for different sample conditions~\footnote{In the course of the experiment, the vacuum chamber had to be vented which
resulted in a shift of eigenfrequencies. The data shown in this manuscript has been obtained after the vent, with the
exception of Fig.~\ref{fig:03NEW}, as the measurement involving the soft ramp was not repeated.}. The extracted splittings for
$m>1$ are obtained with the IAS procedure described previously (see also ~\cite{Chowdhury2022}). For each iteration, we compare
the frequency splitting obtained before (blue datapoints, in the following referred to as unprocessed data) and after applying the
windowing (yellow datapoints, processed data). Notably, the data converges quickly to $\bar{\Omega}_{0}^{(\infty)}/(2\pi) =
(42.65\pm 0.35)\,\mm{kHz}$, obtained after averaging all $\bar\Omega_{0}^{(m)}$ for $m\geq 2$.  Figure~\ref{fig:03} (b) and (c)
display the Ramsey time traces as well as their FFTs for $m=8$. The four Ramsey fringes are clearly visible in the blue trace of
Fig.~\ref{fig:03} (b) before the windowing is applied. Their visibility of approximately $85$ \%  remains unchanged compared to
Fig.~\ref{fig:03NEW} (a). 

The IAS is designed to produce a Ramsey signal with an integer multiple of the oscillation period. Hence, the measured time trace
comprises exactly $4$ fringes, demonstrating that IAS efficiently approaches the real splitting $\Omega_{0}$. The associated FFT
in Fig.~\ref{fig:03} (c) shows a clear maximum at $\SI{44.27\pm 0.08}{\kilo\Hz}$, along with pronounced side lobes caused by
spectral leakage arising from Fourier transforming the short-time data trace without further signal
processing~\cite{Chowdhury2022}).

The yellow trace in Fig.~\ref{fig:03} (b) shows the Ramsey signal after the windowing. The convolution with the window function
effectively halves the number of fringes. This explains why the $n=4$ fringes have to be measured in order to be able to perform a
FFT when using signal processing. The associated FFT signal in Fig.~\ref{fig:03} (c) contains a clear maximum at
$\bar{\Omega}_\text{0}^{(8)}/\lr{2\pi}=\SI{42.79\pm 0.24}{\kilo\Hz}$ and does not exhibit any side lobes. Suppressing spectral
leakage gives rise to a more precise estimate, at the expense of a slightly increased width of the maximum~\cite{Prabhu2018}.
Therefore, the slightly lower value of $\bar{\Omega}_{0}^{(8)}$ extracted from the smaller and broader yellow maximum provides the
better estimate of ${\Omega}_{0}$ than that obtained from the blue trace. Notice that this is also reflected in the size of the
error bars in Fig.~\ref{fig:03} (a). The processed yellow data exhibits larger statistical uncertainties but provides the better
estimate of ${\Omega}_{0}$.

\begin{figure}[t]
    \includegraphics[width=\columnwidth]{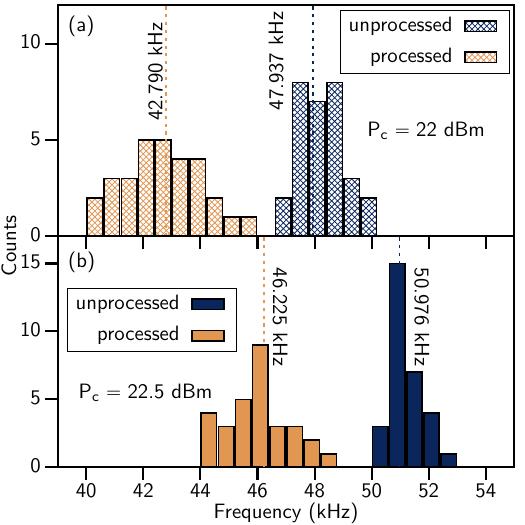}
    \caption{\textbf{Gradient field sensing}. \textbf{(a)} Statistical frequency distributions of $30$ Ramsey measurements at
    iteration $m=8$ for cavity pump power $P_{\mathrm{c}}=\SI{22}{\dBm}$. Blue and orange bars refer to  unprocessed ($n=2$) and
    processed ($n=4$) data, respectively. Dotted lines indicate average $\bar\Omega_0^{(8)}$.  \textbf{(b)} Statistical frequency
    distributions of $30$ Ramsey measurements after increasing the cavity pump power to $P_{\mathrm{c}}=\SI{22.5}{\dBm}$, probed at
    iteration $m=3$.}
    \label{fig:05}
\end{figure}

\textit{Field and charge sensing with IAS ---}
In the following, we will discuss the use of IAS to probe modifications in the size of the frequency splitting ${\Omega}_{0}$
induced by external perturbations. For a proof-of-concept experiment, we perturb the electric field applied between the
electrodes.  The perturbation is induced by increasing the microwave cavity pump power $P_{\mathrm{c}}$ from $\SI{22}{\dBm}$ to
$\SI{22.5}{\dBm}$ which results in a slight shift of the mechanical eigenfrequencies as well as the frequency splitting (see
Appendix~\ref{app:comsol}).  The resulting change of ${\Omega}_{0}$ from the situation discussed in Fig.~\ref{fig:03}
is determined by repeating the IAS measurement as illustrated in Fig.~\ref{fig:05}.   

Figure \ref{fig:05} (a) summarizes the situation before the perturbation is applied, i.e. at $P_{\mathrm{c}} =\SI{22}{\dBm}$. We
show histograms of the statistical frequency distribution of the 30 IAS measurements over $n=4$ fringes at iteration $m=8$  and
indicate their mean value [as also apparent in Fig.~\ref{fig:03} (a)]. The yellow data displays the result of the IAS measurement,
which yields $\bar{\Omega}^{(8)}_\text{0}/\lr{2\pi}=\SI{42.79}{\kilo\Hz}$. For comparison, the blue data illustrates the result of
a regular, un-iterated two-period Ramsey experiment.  As discussed before, the IAS provides the better estimate of $\Omega_{0}$,
even though the data is more spread. Notably, the two histograms are well-separated, providing clear evidence for the failure of
the standard Ramsey experiment for the case of short signals.

Figure \ref{fig:05} (b) displays the frequency estimate obtained after the perturbation, i.e. $P_{\mathrm{c}} =\SI{22.5}{\dBm}$.
According to IAS, here evaluated after iteration $m=3$, the frequency splitting has increased by $\SI{3.435}{\kilo\Hz}$ to
$\bar{\Omega}_\text{0,new}^{(3)}/\lr{2\pi}=\SI{46.23\pm 0.19}{\kilo\Hz}$ (yellow). This corresponds to an added charge density of
$180$\,C/m$^3$ or an equivalent of $1,400$ electrons added to the nanostring (see also Appendix~\ref{app:comsol}). The frequency splitting extracted from the standard, two-period
Ramsey experiment also underwent a frequency shift, but of magnitude $\SI{3.039}{\kilo\Hz}$, approximately $12\,\%$ off from the correct
result. That implies that the two-period Ramsey experiment cannot even be used to extract the relative change of the frequency
splitting induced by the perturbation.
  
\begin{figure}[t]
    \includegraphics[width=\columnwidth]{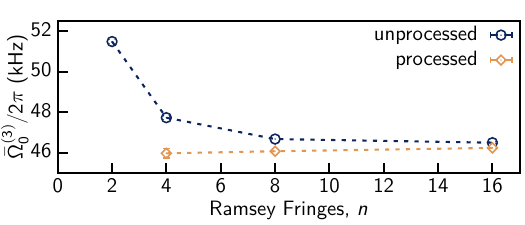}
    \caption{\textbf{Frequency estimate versus number of Ramsey fringes}. Frequency estimation obtained after iteration $m=3$ as a
    function of the number of measured Ramsey fringes $n$ at cavity pump power $P_{\mathrm{c}}=\SI{22.5}{\dBm}$. Blue (orange)
    symbols refer to unprocessed (processed) data, respectively. The unprocessed estimate changes strongly for increasing $n$, whereas
    the processed estimate remains nearly constant. For long signals approximating the standard Ramsey measurement ($n\gg 1$), both
    estimates coincide, validating the IAS approach.}
    \label{fig:04} 
\end{figure}

In order to validate these findings, we repeat the IAS measurement for while varying the number of
Ramsey fringes $n$. The regular Ramsey experiment is expected to converge to the true $\Omega_0$ for $n \to  \infty$.
For the case of IAS, we expect to find the true $\Omega_0$ for a much smaller value of $n$.  Figure~\ref{fig:04} shows
experimental results for $m=3$ iterations. As before, the blue and yellow symbols refer to the unprocessed and processed data,
respectively. As already discussed in Fig.~\ref{fig:03}, the unprocessed data for the shortest time-trace of $n=2$ Ramsey fringes
yields a very inaccurate estimation of $\Omega_0$. For increasing $n$, the accuracy of the estimate increases, as apparent from
the convergence of the extracted frequency near $n=8$. The shortest time-trace suitable for IAS contains $n=4$ Ramsey fringes [see
Fig.~\ref{fig:03} (b)], which already approximates $\Omega_0$ well, within the error margins of the experiment. Not only does this
confirm the reliability of the IAS protocol, it also provides a strong justification for trusting the broad IAS peaks rather than
the more narrow unprocessed results in Figs.~\ref{fig:03} (a) and~\ref{fig:05}. 

\textit{Discussion and outlook ---} 
In conclusion, we experimentally demonstrate high-precision frequency estimation from short Ramsey signals comprising only four
fringes using a nanomechanical two-mode system. This is accomplished by implementing the recently proposed iterative adaptive
spectroscopy (IAS) protocol~\cite{Chowdhury2022}. We show that IAS converges after four iterations, entailing a significant
speedup compared to existing high-precision Ramsey interferometry. In addition, IAS is shown to operate with sparse data sampling
approaching the Nyquist sampling rate.  A proof-of-principle sensing demonstration reveals a change of the coupling strength
subject to perturbations of the electrical field surrounding the nanomechanical string resonator. Exposing the string to an
increased microwave cavity field increases the magnitude of the avoided crossing (see Appendix~\ref{app:comsol}). Notably, IAS allows a
more accurate probe of the resulting change in frequency compared to a standard Ramsey measurement. This holds true both for the
net frequency shift and for the relative change induced by the perturbation, implying that the standard Ramsey method is
inadequate to measure the induced shift thanks to its limitations for short signals. 

We stress that our experimental method is not limited to classical nanomechanical systems but can be applied to any classical or
quantum two-level system~\cite{Taylor2008,Riste2013, Zhou2020,Hollendonner2023,Cheng2023,Cleland2024,Yang2024,Capannelli2025}. Its
speed of operation suggests IAS as a suitable protocol to study the effect of frequency drifts or fluctuations not accessible with
high-resolution Ramsey protocols involving a large number of fringes or even dynamical decoupling strategies. It is equally
well-suited for systems with a short coherence time. 

\textit{Acknowledgements ---} 
We acknowledge funding from the Deutsche Forschungsgemeinschaft (DFG) under Germany’s Excellence Strategy—EXC-2111—390814868.

%


\begin{appendix}
\clearpage
\thispagestyle{empty}
\onecolumngrid
\begin{center}
\textbf{\large Supplemental Material: Precise estimation of the coupling strength between two nanomechanical modes from four Ramsey fringes}\\[1em]

Anh Tuan Le,$^1$ Avishek Chowdhury,$^1$ Hugo Ribeiro,$^2$ and Eva M. Weig$^{1,3,4}$\\[0.5em]

$^1$ \textit{School of Computation, Information and Technology, Technical University of Munich, 85748 Garching, Germany}\\
$^2$ \textit{Department of Physics and Applied Physics, University of Massachusetts Lowell, Lowell, MA 01854, USA}\\
$^3$ \textit{TUM Center for Quantum Engineering (ZQE), 85748 Garching, Germany}\\
$^4$ \textit{Munich Center for Quantum Science and Technology (MCQST), 80799 Munich, Germany}
\end{center}

\setcounter{secnumdepth}{2}
\setcounter{figure}{0}
\renewcommand{\thefigure}{S\arabic{figure}}

\section{Nanomechanical resonator}
\label{app:oop-ip}

The nanomechanical string resonator used in this work is made of strongly pre-stressed stoichiometric silicon nitride
LPCVD-deposited on a fused silica substrate. It has dimensions of $w =250\,\si{\nano\meter}$, $t=100\,\si{\nano\meter}$ and
$L=55\,\si{\micro\meter}$ and is assumed to have simply supported boundary conditions at both clamping points, see
Fig.~\ref{fig:01} (a). 

While its pristine eigenfrequencies can be conveniently obtained from the elastic model of a vibrating string, the situation is
getting more complex once the string is exposed to an electrostatic field that induces dielectric forces~\cite{Rieger2012}.

Finite Element Method (FEM) simulations are an effective tool to study the behaviour of such complex systems. In our work, we use
Comsol Multiphysics to simulate the mechanical properties of the string. This allows to determine the resonant frequency of the
in-plane (IP) and out-of-plane (OOP) flexural modes for different electrostatic field configurations by solving the respective
eigenvalue problem. 

\begin{figure}[h]
	\begin{center}
		\includegraphics[scale=0.7]{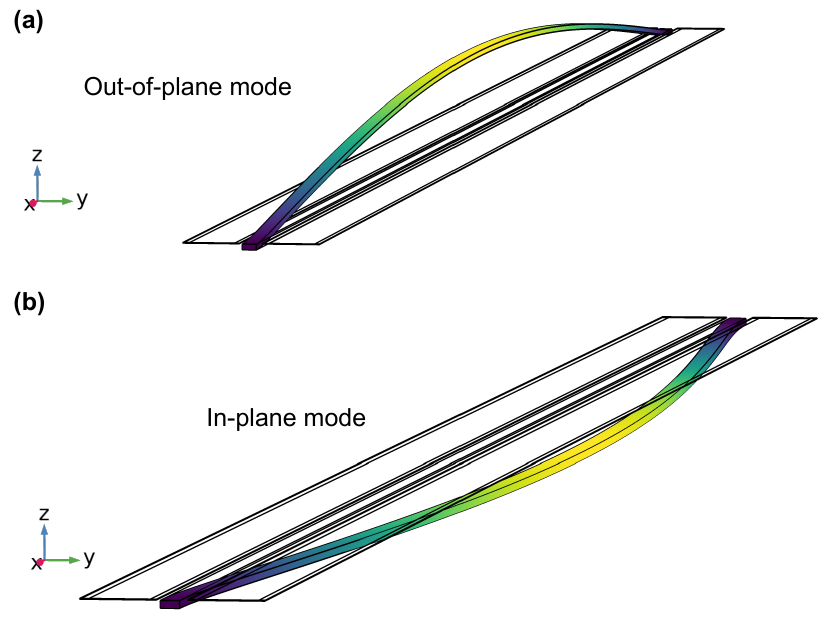}
	\end{center}
    \caption{\textbf{Vibrational modes of a nanomechanical string resonator}. Comsol Multiphysics 3D simulations illustrate the
        fundamental displacement profiles of a doubly clamped nanomechanical string resonator that supports out-of-plane (\textbf{a})
        and in-plane (\textbf{b}) flexural modes.}
    \label{fig:S6}
\end{figure}

In the absence of any applied voltage, our simulation results reveal that the fundamental resonant frequency of the IP mode is
found be to $\omega_{\text{IP}}/\lr{2\pi}=\SI{7.514}{\mega\hertz}$, while the OOP mode is
$\omega_{\text{IP}}/\lr{2\pi}=\SI{7.256}{\mega\hertz}$. The results were obtained through the careful selection of the appropriate
material properties, boundary conditions and mesh refinement.  Note that these frequencies do not exactly coincide with those
measured at a DC voltage of $0$\,V, as a result of the additional dielectric effects of the RMS average of the microwave readout
tone~\cite{Rieger2012}.  The displacement profiles of the IP and OOP flexural modes are illustrated in Figs.~\ref{fig:S6}a
and~\ref{fig:S6}b, respectively.\medskip

\section{Coupled harmonic oscillators}
\label{app:2osc}

The dynamics of the OOP and IP mode follow the equation of motion of two linearly coupled harmonic oscillators
\begin{equation}
    \begin{aligned}
        m\ddot{u}_1 (t) + m \omega_1^2 u_1 (t) + \kappa  \left[ u_1 (t) - u_2 (t)\right] &=
        0, \\
        m\ddot{u}_2 (t) + m \omega_2^2 u_2 (t) + \kappa \left[ u_2 (t) - u_1 (t) \right] & = 0,
    \end{aligned}
\label{eq:couple}
\end{equation}
where $u_1$ and $u_2$ describe the displacement of the OOP and IP mode, respectively~\cite{Novotny2010}. The associated angular
frequencies of the uncoupled modes are denoted by $\omega_j$  ($j=1,2$), and $\kappa$ is the spring coupling constant between the
two modes. In our case, the coupling is dielectrically induced~\cite{Faust2012,barakat2025}.  The effective mass of both modes
$m=m_{\mm{OOP}}=m_{\mm{IP}}=\rho L w t/2$ are equal for a nanostring.  For the sake of simplicity we neglect damping. 

The eigenfrequencies of the two normal modes can be found by substituting the ansatz $u_j (t) = u_{0,j} \exp{(- i \omega t)}$ in
Eq.~\eqref{eq:couple}. This leads to a system of linear equations for the amplitudes $u_{0,j}$ ($j=1,2$) that admit non-trivial
solution when $\omega = \omega_\pm$. We find  
\begin{equation}
    \omega_{\pm}^2= \frac{\omega_1^2+\omega_2^2 + 2 \omega_\kappa^2 \pm\sqrt{(\omega_1^2-\omega_2^2)^2+4 \omega_\kappa^4}}{2},
\label{eq:freqs}
\end{equation}
where we have defined $\omega_\kappa = \sqrt{\kappa/m}$. From Eq.~\eqref{eq:freqs}, we can find the minimal level splitting $\Omega_0$, 
which corresponds to the frequency splitting when $\omega_1 = \omega_2$. We have 
\begin{equation}
    \omega_+^2 - \omega_-^2= (\omega_+ - \omega_-)(\omega_+ + \omega_-) =\sqrt{(\omega_1^2-\omega_2^2)^2 + 4 \omega_\kappa^4},
\end{equation}
which leads to 
\begin{equation}
   \omega_+ - \omega_- =  \frac{\sqrt{(\omega_1^2-\omega_2^2)^2 + 4 \omega_\kappa^4}}{\omega_+ + \omega_-}.
    \label{eq:DiffFreqExact}
\end{equation}
Assuming $k_1, k_2 \gg \kappa$, and setting $\omega_1 = \omega_2$ in Eq.~\eqref{eq:DiffFreqExact}, we find that the minimal level splitting is given by 
\begin{equation}
    \Omega_0 \simeq \frac{\omega_k^2}{\omega_1} = \frac{\omega_k^2}{\sqrt{\omega_1 \omega_2}}.
    \label{eq:Omega0}
\end{equation}

\begin{figure}[h]
	\begin{center}
		\includegraphics[scale=1]{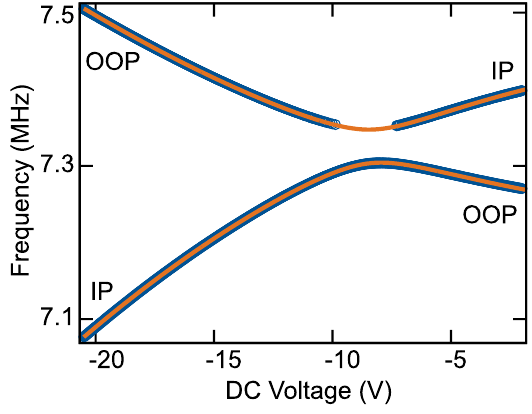}
	\end{center}
    \caption{\textbf{Avoided Crossing}. Dielectric frequency tuning of the in-plane (IP) and out-of-plane (OOP) mode of the
        nanostring resonator. Blue circles indicate the eigenfrequencies obtained from Lorentzian fits of the spectroscopic
        measurement depicted in Fig.~\ref{fig:01} (c) of the main text. Orange lines show the fit using Eq.~\eqref{eq:freqs}.}
    \label{fig:S3}
\end{figure}

Substituting Eq.~\eqref{eq:Omega0} in Eq.~\eqref{eq:freqs} allows us to express $\omega_\pm$ as a function of the measurable frequency $\Omega_0$. We find
\begin{equation}
    \omega_{\pm}^2= \frac{\omega_1^2+\omega_2^2 + 2 \Omega_0 \sqrt{\omega_1 \omega_2} \pm\sqrt{(\omega_1^2-\omega_2^2)^2+4 \Omega_0^2 \omega_1\omega_2}}{2}.
    \label{eq:freqsOmega0}
\end{equation}

Changing the DC voltage applied between the electrodes allows tuning the eigenfrequencies~\cite{Rieger2012}. This allows one to
map out the avoided crossing between the modes, see Fig.~\ref{fig:S3}.  Note the missing data in the upper branch in the region of
the avoided crossing. The low visibility in the spectroscopic measurement [see Fig.~\ref{fig:01} (c)] hinders the direct
extraction of the respective eigenfrequencies.  Fitting the above model [see Eq.~\eqref{eq:freqsOmega0}] to the data (orange line)
reveals a splitting of $\SI{41.3}{\kilo\hertz}$. This value is used as the prior estimate for $\Omega_{0}/(2\pi)$.

 \section{3D microwave cavity-assisted detection scheme for nanomechanical string resonators}
 \label{app:3Dcavity}

\begin{figure}[h]
	\begin{center}
		\includegraphics[scale=.9]{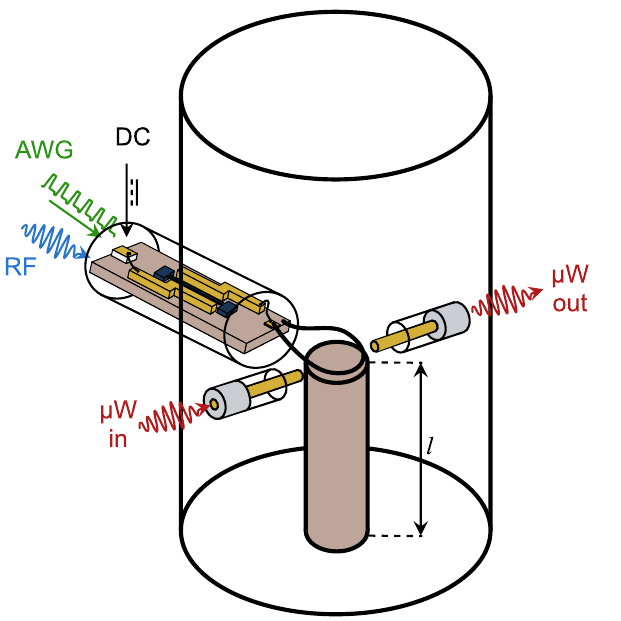}
	\end{center}
    \caption{\textbf{Coaxial $\lambda /4$ microwave cavity}. Schematic of the 3D microwave cavity and coupling architecture used
        for dielectric displacement detection of the nanostring.}
    \label{fig:S1}
\end{figure}

All measurements described in this work have been conducted using the 3D microwave cavity-assisted detection scheme illustrated in
Fig.~\ref{fig:S1}. It relies on a coaxial $\lambda/4$ microwave cavity~\cite{Reagor2013}. This type of cavity consists of a narrow
cylindrical waveguide with an inner conductor of length $l$ that is shorted-circuited on one end and open on the other.  To date,
this type of cavity has only been realized at cryogenic temperatures for circuit cavity electrodynamics with superconducting
circuits~\cite{Reagor2013}, while other types of superconducting 3D cavities have already been adapted for cavity
electromechanics~\cite{Yuan2015a, Noguchi2016,Carvalho2019}.  Here we use a coaxial $\lambda/4$ cavity made of copper, suitable
for room-temperature operation. Thanks to its clean, single mode-like spectrum and its large quality factor, it is better suited
for nanomechanical displacement detection than the planar PCB-based~\cite{Faust2013} or 3D cylindrical cavity~\cite{Le2021}
employed in previous room-temperature experiments.  Our cavity exhibits a waveguide radius and length of $8$\,mm and $40$\,mm,
respectively, and an inner conductor of radius and length of $1.8$\,mm and $l=18$\,mm, respectively. The eigenfrequency of the
fundamental TEM mode $\omega_c$ is defined by the length of the inner conductor $l\approx \lambda/4$. In our case, we find
$\omega_c/(2\pi) \approx 3.7$\,GHz for the bare cavity, and $\omega_c/(2\pi) \approx 3.0$\,GHz for the cavity loaded with the
antenna and the sample holder. All other modes of the cavity, including the second harmonic TEM mode as well as the cylindrical
waveguide mode TE$_{11}$, are found at frequencies above $10$\,GHz.  The cavity is pumped on resonance to avoid effects of
dynamical backaction on the nanostring resonator. 

The nanomechanical string resonator is coupled to the cavity via a microwave antenna that is wire-bonded to one of the gold
electrodes. The second gold electrode serves as a microwave ground (see Appendix~\ref{app:setup}). The presence of the antenna and the
sample holder leads to a reduction of the cavity eigenfrequency to $\omega_c \approx 3$\,GHz.  As described in the main text, the
mechanical motion at eigenfrequency $\omega_j$ ($j\in \{\mm{OOP}, \mm{IP}\}$) periodically modulates the capacitance of the
electrodes. For the resonantly pumped cavity, this induces sidebands in the cavity response at frequency $\omega_c\pm\omega_j$,
which are recorded via the cavity transmission. 

The dynamics of the string is controlled via the dielectric drive and control signal, which is also applied via the second gold
electrode. It consists of a drive tone (RF), an arbitrary waveform to implement the corrected ramp required for IAS (AWG), as well
as the DC voltage. More details about the dielectric drive and control scheme, as well as the demodulation about the microwave
signal can be found in Appendix~\ref{app:setup} as well as in Fig.~\ref{fig:01} (b) of main text.

\section{Experimental Setup}
\label{app:setup}
 
\begin{figure}[h]
	\begin{center}
		\includegraphics[scale=0.65]{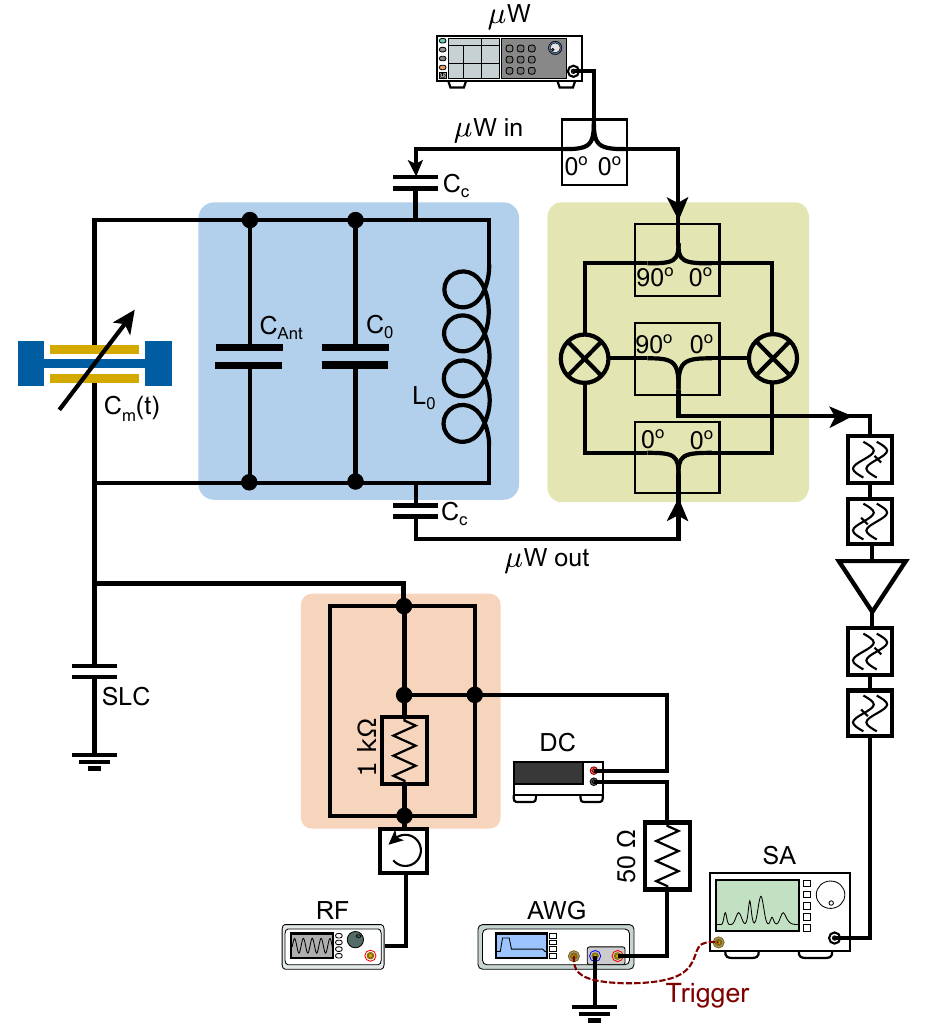}
	\end{center}
    \caption{\textbf{Circuit diagram of the measurement setup.} The resonator (blue) with its adjacent electrodes (yellow) are
    connected to the microwave cavity and the IQ-mixer (shaded in blue and green, respectively) for readout, and to the voltage
    combiner box (shaded in red) merging the input signals. The single layer capacitor (SLC) provides a capacitive ground for the
    microwave signal.}
    \label{fig:S2}
\end{figure}

Figure~\ref{fig:S2} illustrates the full circuit diagram of the experimental setup. A microwave drive tone from the signal
generator ($\mu$W, R\&S SMB100) is split into two arms. The first arm pumps the 3D microwave coaxial $\lambda/4$ cavity on
resonance at frequency $\omega_c/\lr{2\pi}\approx\SI{3}{\giga\hertz}$ to avoid optomechanical backaction effects. The cavity is
represented by its $L$-$C$ equivalent circuit along with the antenna capacitance $C_\mm{Ant}$ (blue-shaded box) coupling the
cavity to one of the gold electrodes as described in Appendix~\ref{app:3Dcavity}. The second gold electrode provides a ground for the
microwave signal via a single layer capacitor (SLC). The mechanical motion of the nanostring periodically modulates the
capacitance $C_m(t)$ between the electrodes. This results in a phase modulation of the cavity output signal and imprints sidebands
at $\omega_c\pm\omega_j$ onto the cavity response. These are demodulated by mixing the cavity transmission with the reference arm
using an IQ-mixer (green-shaded box). The demodulated quadratures are subsequently combined using a $0^{\circ}/90^{\circ}$ power
splitter~\cite{Faust2012} (also included in green-shaded box). After filtering out the frequency component at $\omega_j$ and
amplification of the signal, the amplitude and phase of the mechanical signal is recorded with a spectrum analyzer (SA, R\&S FSV).
The arbitrary waveform generator (AWG, Keysight 81150A) provides a trigger signal to start the measurements. All devices are
synchronized to its internal $\SI{10}{\mega\hertz}$ clock.
 
Dielectric actuation and control of the nanomechanical resonator is achieved by applying a voltage control signal to the second
gold electrode via the SLC~\cite{Rieger2012, Unterreithmeier2009}. 

 \begin{itemize}
     \item For the case of a simple spectroscopic characterization as in Fig.~\ref{fig:01} (c), the voltage control signal
         consists of a DC voltage and an RF drive tone. These signals can be conveniently added using a conventional bias tee.
    \item For the case of IAS, the voltage control signal comprises the DC voltage (DC Source, Keithley 2410), an RF drive tone
        from a function generator in the MHz regime (RF, Keysight 33500B), and an arbitrary waveform (AWG, Keysight 81150A)
        providing the voltage ramps. The latter consists of multiple frequency components in the kHz range. This signal cannot be
        added with a regular bias tee, as its components neither obey the high-frequency cutoff of its DC, nor the low-frequency
        cutoff of the RF path.  To avoid unfaithful pulse generation resulting from bandwidth limitations of the bias tee, we
        chose a different voltage combination approach based on a series connection of the DC voltage sources and the ramp signal
        provided by the AWG.

        This is implemented in a home-built voltage combiner box (red-shaded box), shown in more detail in Fig.~\ref{fig:S5}~(a).
        The high impedance nanomechanical string resonator flanked by the two gold electrodes is represented by the circuit
        elements in the green dashed box. The AWG ramp signal $V_1$ and the DC voltage $V_3$ are connected in series.  Proper
        grounding through the box is ensured by floating the DC source.  

        The RF drive tone $V_2$ from the function generator is added to the DC-offset AWG signal via a $1$\,k$\Omega$ impedance.
        This results in a strong attenuation of $V_2$, ensuring operation of the nanostring in the linear response regime, and
        also provides some isolation.  

        In Fig.~\ref{fig:S5} (b), we plot the transmission of the AWG's output voltage $V_1$ as a function of frequency in a Bode
        diagram. At low frequencies, a constant transmission close to $0$\,dB is obtained. The near-negligible attenuation of
        approximately $-0.4$\,dB results from the minimal loading thanks to the large difference in the output impedances in the two
        paths ($50\,\Omega$ vs. $1\,$k$\Omega$).  The transmission starts decreasing at frequencies exceeding $100$\,kHz. This
        shows that the voltage combiner box allows adding AWG signals with a bandwidth of up to approximately $100$\,kHz, a significant
        improvement compared to a standard bias tee.

\begin{figure}[h]
	\begin{center}
		\includegraphics[scale=1]{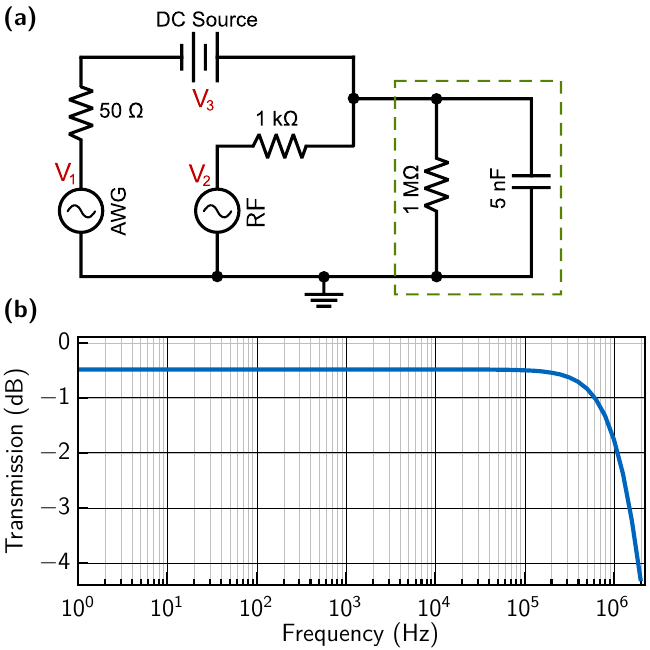}
	\end{center}
	\caption{\textbf{Voltage combiner box.} (\textbf{a}) Circuit diagram. (\textbf{b}) Bode diagram.}
\label{fig:S5}
\end{figure}

In the present work, two different types of voltage ramps are employed:
\begin{itemize}[label=$\circ$]
    \item The soft ramp is implemented demanding $\Omega_0t_s/\lr{2\pi} = 1/2$, as discussed in detail in
        Ref.~\cite{Chowdhury2022}. This corresponds to a sinusoidal voltage signal oscillating at $20.8$\,kHz (corresponding to
        $t_\us = 12\,\mu$s), well within the bandwidth of the voltage combiner box.
    \item IAS relies on corrected ramps based on the Magnus-based strategy for control. These  require higher order frequency
        components (see Appendix~\ref{app:pulse}). For example, the first harmonic Magnus-based correction used in the present work
        [c.f. Fig.~\ref{fig:02} (a)] combines a $20.8$\,kHz and a $41.6$\,kHz signal, equally well within the bandwidth of the
        box.

        Notice that higher harmonic Magnus-based corrections involving additional, higher-frequency components, can, in principle,
        increase the accuracy of the state generation. However, they will get close to or exceed the bandwidth of the voltage combiner
        box and therefore lead to an unfaithful pulse shape on the sample, and have not been explored in the context of this work.
    \end{itemize}
 \end{itemize}

\section{Ramsey protocol}
\label{app:ramsey}

The Ramsey protocol is a spectroscopic method that allows for the resolution of the frequency splitting between two states. In the
following, we will use the ket notation to label the two classical states as $\ket{0}$ and $\ket{1}$. These states are
conveniently visualized on a Bloch sphere.  In general, the Ramsey sequence consists of a series of five distinct steps:
\begin{enumerate}[label=(\Roman*)]
    \item The two-level system is prepared in one of the chosen basis states $\ket{0}$ or $\ket{1}$ as a fiducial state.
    \item A equally-weighted superposition state of the two basis states $\ket{\psi} = (\ket{0} - e^{i\varphi}\ket{1})/\sqrt{2}$
        with an arbitrary phase $\varphi$ is prepared as the sensing state. 
    \item By letting $\ket{\psi}$ evolve freely for a time $t_\uw$, the relative phase between $\ket{0}$ and $\ket{1}$ is
        evolving, giving rise to a precession of the state in the superposition plane of the Bloch sphere.  
    \item The final state $\ket{\psi (t_\uw)}$ is projected back into one of the two original basis states as the readout state.
    \item The return probability of finding the readout state is measured as a function of $t_\uw$. 
\end{enumerate}

\begin{figure}[h!]
	\begin{center}
		\includegraphics[width=0.99\columnwidth]{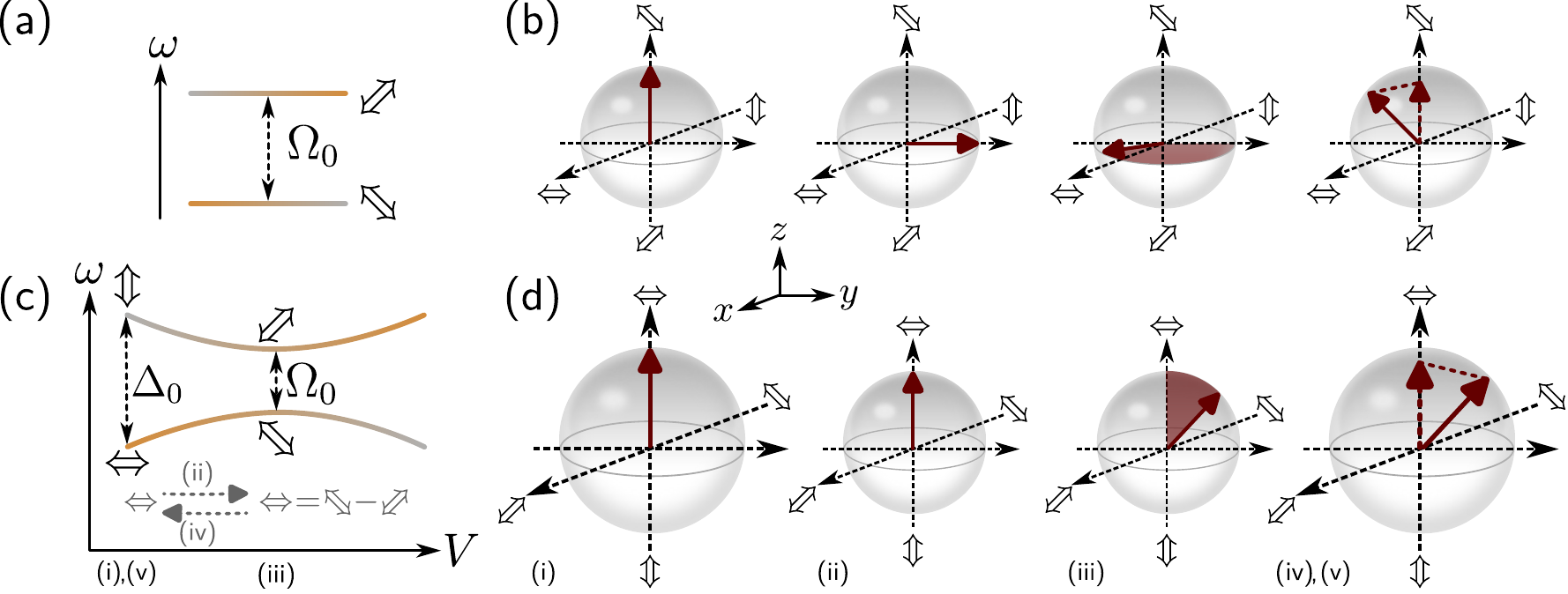}
	\end{center}
    \caption{\textbf{Comparison of Ramsey protocols.} (\textbf{a}), (\textbf{b}) Ramsey protocol in the basis of normal modes
        $\np$ and $\nm$. (\textbf{a}) Splitting between the normal modes $\np$ and $\nm$ at the avoided crossing. (\textbf{b})
        Definition of Bloch sphere and Ramsey sequence in the basis defined in (\textbf{a}).  (\textbf{c}), (\textbf{d}) Ramsey protocol
        in the basis of uncoupled modes $\Updownarrow$ and $\Leftrightarrow$. (\textbf{c}) Avoided crossing in the basis uncoupled modes
        $\Updownarrow$ and $\Leftrightarrow$, hybridizing into normal modes $\np$ and $\nm$ on resonance. (\textbf{d}) Definition of Bloch
        sphere and Ramsey sequence in the basis defined in (\textbf{c}). Lower case Roman numerals in (\textbf{c}) and (\textbf{d})
        indicate steps of the Ramsey sequence described in the main text.}
    \label{fig:Blochsphere}
\end{figure}

There are several ways to physically assign the basis states $\ket{0}$ and $\ket{1}$. This is illustrated in
Fig.~\ref{fig:Blochsphere}, which describes two distinct ways of implementing the Ramsey sequence on a nanomechanical two-mode
system.

Figure~\ref{fig:Blochsphere} (a) depicts the energy spectrum of a two-mode system with splitting $\Omega_0$. It can be thought of
as the two nanomechanical modes tuned on resonance (with hybrid normal modes $\np$ and $\nm$), but also corresponds to the
standard configuration of a spin $1/2$ or a qubit. In this case, the normal modes $\np$ and $\nm$ are chosen as the basis states.
This gives rise to the Ramsey sequence illustrated in Fig.~\ref{fig:Blochsphere} (b), which corresponds to the conventional Ramsey
sequence~\cite{Degen2017}. 

Its five distinct steps are implemented as follows:
\begin{enumerate}[label=(\Roman*)]
    \item The two-level system is prepared in its ground state $\nm$, as depicted in the first Bloch sphere of
        Fig.~\ref{fig:Blochsphere} (b). For a nanomechanical two-mode system this can be accomplished by slow adiabatic passage
        which brings the system from one of its uncoupled eigenmodes $\Updownarrow$ or $\Leftrightarrow$ into the avoided level
        crossing~\cite{Faust2013}.  
    \item The equally-weighted superposition state (sensing state) is generated by applying a resonant $\pi/2$ pulse. It rotates
        the state into the equatorial plane of the Bloch sphere, as illustrated in the second Bloch sphere of
        Fig.~\ref{fig:Blochsphere} (b).
    \item The sensing state precesses in the equatorial plane during time $t_\uw$. Notice that for the case of the nanomechanical
        two-mode system, it will oscillate between the original IP and OOP mode, see third Bloch sphere of
        Fig.~\ref{fig:Blochsphere} (b).
    \item A second $\pi/2$ pulse is used to project the final state back into the readout state $\nm$, as shown in the fourth
        Bloch sphere of Fig.~\ref{fig:Blochsphere} (b).
    \item The probability of finding the readout state $\nm$ is measured. This is also illustrated in the fourth Bloch sphere of
        Fig.~\ref{fig:Blochsphere} (b). For a nanomechanical two-mode system, this is done using a second, reverse adiabatic
        passage followed by a ringdown measurement of an uncoupled eigenstate ($\Updownarrow$ or
        $\Leftrightarrow$)~\cite{Faust2013}. 
\end{enumerate}

Figure~\ref{fig:Blochsphere} (c) shows the energy spectrum of a tunable two-mode system. It exhibits an avoided level crossing
with splitting $\Omega_0$ which can be addressed by tuning the voltage $V$. Far from the avoided crossing, the system is best
described by its uncoupled eigenstates $\Updownarrow$ and $\Leftrightarrow$. This is the case as long as their frequency
separation greatly exceeds $\Omega_0$. The Ramsey sequence illustrated in Fig.~\ref{fig:Blochsphere} (d) starts at a particular
initialization point where this separation is $\Delta_0$. For the nanomechanical two-mode system the uncoupled eigenstates
correspond to the OOP and IP mode, respectively. When the system is tuned towards the avoided crossing, the uncoupled eigenstates
start hybridizing into the normal modes $\nm$ and $\np$. The small Roman numberals indicate the steps of the Ramsey sequence
illustrated in Fig.~\ref{fig:Blochsphere} (d) and described in the main text. This Ramsey sequence is best described using the
basis of the uncoupled eigenstates $\Updownarrow$, $\Leftrightarrow$, which map onto the south and north pole of the Bloch sphere
used in Fig.~\ref{fig:Blochsphere} (d). The equatorial plane hosts equal superpositions of the two basis states. On the x-axis of
the Bloch sphere, we find $\np = (\Updownarrow + \Leftrightarrow)/\sqrt{2}$ and $\nm = (\Updownarrow - \Leftrightarrow)/\sqrt{2}$,
recovering the normal modes. This choice of basis gives rise to the Ramsey sequence illustrated in Fig.~\ref{fig:Blochsphere} (d). 

Its five distinct steps are implemented as follows (see also~\cite{Chowdhury2022} for more details):
\begin{enumerate}[label=(\roman*)]
    \item The initialization of the system is performed far-detuned from the avoided level crossing [c.f. label (i) in
        Fig.~\ref{fig:Blochsphere} (c)]. In this configuration, we chose the IP mode $\Leftrightarrow$ as the initialization
        state. The energy separation of the two eigenmodes at the initialization point corresponds to $\Delta_0 \ll \Omega_0$.
        This is reflected in a larger radius of the first Bloch sphere of Fig.~\ref{fig:Blochsphere} (d). 
    \item To prepare the sensing state, the system is brought into the avoided level crossing using a voltage sweep [c.f. label
        (ii) in Fig.~\ref{fig:Blochsphere} (c)]. The sensing state represents an equally-weighted superposition state of the two
        eigenmodes at the avoided crossing, i.e., the normal modes $\np$ and $\nm$. Expressed in the chosen basis, they correspond
        to $\Leftrightarrow =(\np - \nm)/\sqrt{2}$ and $\Updownarrow =(\np + \nm)/\sqrt{2}$. So unlike in the conventional Ramsey
        configuration, the sensing state is oriented \emph{ along} the z-axis of the Bloch sphere. Here we chose $\Leftrightarrow$
        as the sensing state.  The choice of initialization and sensing state (both $\Leftrightarrow$, but at two different bias
        points, reflected in a modified radius of the Bloch spheres) implies that the voltage sweep must generate an operation
        proportional to an identity. In principle, this can be realized by an instantaneous voltage sweep. In practice, given the
        bandwidth limitations of arbitrary waveform generators, such a sweep is impractical. Instead, one can resort to control
        techniques to realize the desired dynamics. Here, we use the Magnus-based strategy for control to generate an identity
        operation, see Appendix~\ref{app:pulse} for more details.
    \item The sensing state precesses in the superposition plane of the normal modes $\np$ and $\nm$. Here this corresponds to the
        $y$-$z$-plane. The state evolves freely during time $t_\uw$.
    \item A reverse voltage sweep is used to project the final state back into the far-detuned readout state $\Leftrightarrow$
        [c.f. label (iv) in Fig.~\ref{fig:Blochsphere} (c)]. As discussed before, this requires an identity operation. The change
        in bias point is illustrated in a modified radius of the Bloch sphere. We use the time-reverse of the pulse described in
        step (ii), c.f. Fig.~\ref{fig:02} (a).
    \item The probability of finding the far-detuned readout state $\Leftrightarrow$ is measured. 
\end{enumerate}

The two Ramsey sequences described in Figs.~\ref{fig:Blochsphere} (b) and (d) both yield a signal which contains spectral
information about the splitting $\Omega_0$. However, the different choice of basis has certain implications: From a mathematical
point of view, the signal in the conventional Ramsey sequence, c.f. Figs.~\ref{fig:Blochsphere} (b), is equivalent to measuring
the expectation value of the Pauli matrix $\sigma_z = \ketbra{\np}{\np}-\ketbra{\nm}{\nm}$, whereas the signal in the Ramsey
sequence implemented in the present work, c.f. Figs.~\ref{fig:Blochsphere} (d), is equivalent to measuring the expectation value
of the Pauli matrix $\sigma_z = \ketbra{\Updownarrow}{\Updownarrow}-\ketbra{\Leftrightarrow}{\Leftrightarrow}$. The two different
choices of $\sigma_z$ are related by a rotation of the Bloch sphere as detailed in the following: The Bloch sphere of interest is
obtained from the conventional Bloch sphere by a $-\pi/2$ rotation of the Bloch sphere in Fig.~\ref{fig:Blochsphere} (b) around
its y-axis. Correspondingly, the conventional Bloch sphere is recovered from the Bloch sphere in Fig.~\ref{fig:Blochsphere} (d) by
a $\pi/2$ rotation around its y-axis.  In other words, using the choice of basis and definition of Fig.~\ref{fig:Blochsphere} (b),
the Ramsey measurement of Fig.~\ref{fig:Blochsphere} (d) is equivalent to a measurement of $\sigma_x$ [in the basis of the hybrid
modes $\nm$, $\np$ used in Figs.~\ref{fig:Blochsphere} (b)].

For the nanomechanical two-mode system (and for \emph{any} other tunable two-mode system) one can choose to implement either of
the two Ramsey sequences. For example, the conventional Ramsey sequence [Figs.~\ref{fig:Blochsphere} (a) and (b)] was used to
demonstrate full Bloch sphere control and Stückelberg interferometry~\cite{Faust2013,Seitner2016,Seitner2017}. Here, we opt for
the other Ramsey sequence [Figs.~\ref{fig:Blochsphere} (c) and (d)], which allows for a faster initialization. Given that the
nanomechanical two-mode system needs to be initialized far from the avoided level crossing to avoid spurious effects of the drive,
the implementation of the conventional Ramsey sequence requires a slow adiabatic passage to bring the system from the initialized
uncoupled eigenmode ($\Updownarrow$ or $\Leftrightarrow$) into the avoided level crossing. On the other hand, as detailed in the
main text as well as in Appendix~\ref{app:pulse}, the initialization for the Ramsey sequence under investigation calls for a fast,
non-adiabatic pulse. 
 
\section{Nanomechanical implementation of IAS}
\label{app:pulse}

The dynamics of the IAS protocol is modeled by the Hamiltonian~\cite{Chowdhury2022}
\begin{equation}
    H(t)=\frac{\Delta(t)}{2}\sigma_z+\frac{\Omega_0}{2}\sigma_x,
    \label{eq:dynamicalmatrix}
\end{equation}
where $\sigma_x$, $\sigma_y$ and $\sigma_z$ are the Pauli matrices [with $\sigma_x =$ $\ketbra{\Updownarrow}{\Leftrightarrow}$
$+\ketbra{\Leftrightarrow}{\Updownarrow}$, $\sigma_y =$ $ -i \ketbra{\Updownarrow}{\Leftrightarrow}$
$+i\ketbra{\Leftrightarrow}{\Updownarrow}$, $\sigma_z =$ $\ketbra{\Updownarrow}{\Updownarrow}$
$-\ketbra{\Leftrightarrow}{\Leftrightarrow}$, c.f. Appendix~\ref{app:ramsey} and Figs.~\ref{fig:Blochsphere} (c) and (d)] and
$\Delta(t)=\omega_\text{OOP}(t)-\omega_\text{IP}(t)$ is the frequency difference between the OOP and the IP mode. This quantity is
tunable via the applied voltage $U(t)$. For a simple spectroscopic measurement as in Fig.~\ref{fig:01} (c), this is implemented by
a DC voltage sweep~\cite{Rieger2012,Unterreithmeier2009}, while for the more complex IAS sequence the arbitrary waveform generator
is used to generate $\Delta [U(t)]$ (see Appendix~\ref{app:setup}).

The voltage pulse $U(t)$ is defined according to the steps (i) - (v) of the sequence defined in the main text [see also
Fig.~\ref{fig:02}(a)] as
\begin{equation}
	U (t) =
	\begin{cases}
		U_i, &\text{for}\ t < t_0, \\
		U_i + \Delta U\cdot g_\mm{lead}(t), &\text{for}\ t_0 \leq t < t_s, \\
		U_f,  &\text{for}\ t_s \leq t < t_f,\\
		U_f - \Delta \bar{U}\cdot g_\mm{trail} (t), &\text{for}\ t_f \leq t < t_r,\\
		U_r, &\text{for}\ t \geq t_r. \\
	\end{cases}
	\label{eq:RamseyScheme}
\end{equation}
Here, $U_i$, $U_f$, and $U_r$ are the initial, final, and readout voltages, respectively [see Fig.~\ref{fig:02} (a)]. The voltage
differences $\Delta U=U_f-U_i$ and $\Delta \bar{U} = U_f - U_r$ are chosen to differ to allow the readout mode to decay for the
ringdown measurement (see main text for more details). The times $t_0$, $t_s$, $t_f$, and $t_r$ are chosen such that $t_s - t_0 =
t_r - t_f$.

To find a non-adiabatic frequency-sweep that prepares with high-fidelity the sensing state, we rely on the Magnus-based strategy
for control~\cite{Ribeiro2017,Roque2021} as described in Ref.~\cite{Chowdhury2022} (and Supplemental Material section III of
thereof).  The first step consists in choosing a first frequency-sweep (the soft ramp introduced in the main text), and which will
then be modified by a correcting pulse. The generic form of the corrected ramp is then given by 
\begin{equation}
    g_a (t)=g_{a,\mm{soft}}(t) + g_{a,\text{corr}}(t),
    \label{eq:CorrRamp}
\end{equation}
with $a \in \{\mm{lead},\,\mm{trail}\}$. Here, we choose the leading edge of the soft ramp to be given by [see Fig.~\ref{fig:02}
(a) of the main text]
\begin{equation}
    g_{\mm{lead},\mm{soft}}(t) = \frac{1}{2}\left[1-\cos\left(\pi\frac{t-t_0}{t_s-t_0}\right)\right]
    \label{eq:LeadSoftRamp}
\end{equation}
and the correcting sweep is parametrized as 
\begin{equation}
    g_{\mm{lead},\text{corr}}(t)=g_{\mm{lead},\text{even}}(t) + g_{\mm{lead},\text{odd}}(t)
    \label{eq:MagnusSweep}
\end{equation}
with 
\begin{equation}
    \begin{aligned}
        g_{\mm{lead},\text{even}}(t)& =c \left[1-\cos\left(2\pi\frac{t-t_0}{t_s-t_0}\right)\right],\\
        g_{\mm{lead},\text{odd}}(t)& = d \left[\sin\left(2\pi\frac{t-t_0}{t_s-t_0}\right)\right],
    \end{aligned}
    \label{eq:magnusterms}
\end{equation}
where $c$ and $d$ are the free Fourier coefficients that allow one to tune the correcting sweep to cancel transitions to the
out-of-plane (OOP) mode. These coefficients are found using the Magnus-based strategy for control as described in the Supplemental
Material of Ref.~\cite{Chowdhury2022}.

We can relate the trailing edge to the leading edge of the non-adiabatic detuning-sweep via the relation 
\begin{equation}
    g_\mm{trail,soft} (t) = 1 - g_{\mm{lead},\mm{soft}}(t),
    \label{eq:TrialEdge}
\end{equation}
which we then use to find the coefficients $\bar{c}$ and $\bar{d}$ for the trailing edge correcting sweep [see
Eqs.~\eqref{eq:RamseyScheme} and \eqref{eq:magnusterms}]. Since $\Delta U \neq \Delta \bar{U}$ there is no symmetry relation
between the coefficients $c$ ($d$) and $\bar{c}$ ($\bar{d}$). Note that this arises from the technical implementation of the
experiment as discussed in the main text, and it is not a limitation of IAS~\cite{Chowdhury2022}.

\section{Nanomechanical string resonator as a sensor}
\label{app:comsol}

To validate the sensing capability of the nanomechanical string resonator, we perform finite element simulations using COMSOL
Multiphysics to study the effect of electrostatic fields on the mechanical eigenfrequencies. To this end, we add two adjacent
electrodes to the model of the string described in Sec.~\ref{app:oop-ip}. To reproduce the dielectric tuning of our
sample~\cite{Faust2012a} [c.f. Figs.~\ref{fig:01} (c) and~\ref{fig:S3}], we mimick the nominal layout of the sample summarized in
Tab.~\ref{tab:parameters}. 

\begin{table}[h!]
  \centering
  \begin{tabular}{lll}
\arrayrulecolor{MidnightBlue}
    \bottomrule
    Parameter & Symbol & Current device 
    \\
    \hline
    Length & $L$ & $\SI{55}{\micro\meter}$  \\
    Width & $w$ & $\SI{250}{\nano\meter}$ \\
    Thickness & $t$ & $\SI{100}{\nano\meter}$  \\
  
    Electrode width & W  & $1\,\mu$m \\
    Electrode separation& S  & $\lesssim 500$\,nm    \\
    Electrode asymmetry & A  & unknown    \\
    \bottomrule
  \end{tabular}
    \caption{\textbf{Geometric parameters of the nanostring resonator.} The true electrode geometry, especially the gap between
    the two electrodes $S$ as well as the exact position of the nanostring determining the asymmetry of the capacitor $A$, is not
    precisely known.}
  \label{tab:parameters}
\end{table}

\begin{figure}[h]
	\begin{center}
		\includegraphics[scale=1]{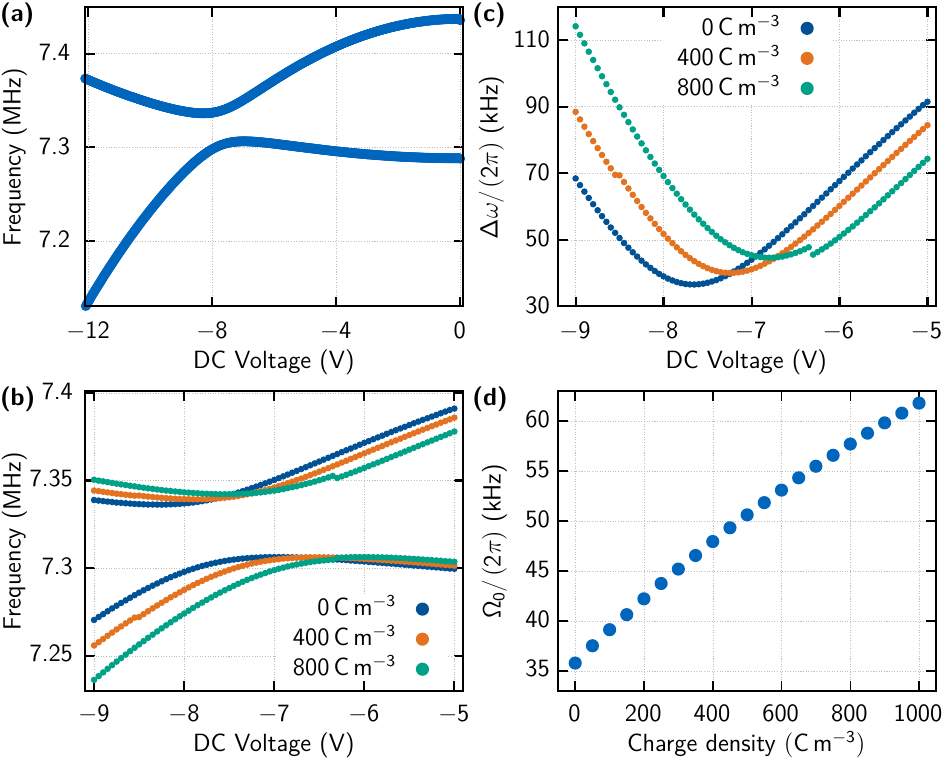}
	\end{center}
    \caption{\textbf{COMSOL Multiphysics simulation of the dielectric eigenfrequency tuning of the nanomechanical string resonator}. 
        (\textbf{a}) Eigenfrequency tuning of the fundamental IP and OOP flexural mode as a function of an applied DC bias
    voltage. (\textbf{b}) Closeup in the vicinity of the avoided crossing for three different charge densities on the nanostring,
    mimicking different RMS microwave fields ($0$\,C$/$m$^{-3}$, $400$\,C$/$m$^{-3}$, $800$\,C$/$m$^{-3}$ in blue, orange and green,
    respectively). (\textbf{c}) Frequency difference $\Delta\omega$ between the two branches in (b) for the same three charge
    densities.  (\textbf{d}) Level splitting $\Omega_0$ corresponding to the minimum in (c) as a function of the charge density in the
    string.}
    \label{fig:S7}
\end{figure}

We apply a DC bias between the electrode to expose the string to an electostatic gradient field and perform an eigenfrequency
analysis to assess the dielectric frequency shift of the OOP and IP modes. Figure~\ref{fig:S7} (a) shows the tuning behavior of
both modes as the DC bias is swept from $0$\,V to $-12$\,V. The quadratic tuning behavior of both modes is apparent, reproducing
the expected softening (stiffening) behavior of the IP (OOP) mode, respectively~\cite{Faust2012a,Rieger2012}. Close to $V_\mm{DC}
= -8\,\mm{V}$ the avoided crossing clearly emerges as a result of the dielectric coupling between the two modes~\cite{Faust2012a},
yielding a splitting $\Omega_0/(2\pi) = 36$\,kHz.  

Besides the applied DC voltage, the microwave tone applied to the 3D microwave cavity for readout (c.f. Appendix.~\ref{app:3Dcavity})
also induces an effective DC voltage, arising from the root-mean-square (RMS) of the microwave field~\cite{Rieger2012}. We also
account for the RMS microwave field in the COMSOL simulation. For the sake of simplicity, we describe its effect by adding a net
charge density to the dielectric material of the beam, rather than adding the microwave cavity to the model. Figure~\ref{fig:S7}
(b) displays the original DC bias sweep from Fig.~\ref{fig:S7} (a) (charge density $0$\,C$/$m$^{-3}$, blue trace) along with two
other charge configurations (charge density $400$\,C$/$m$^{-3}$ and $800$\,C$/$m$^{-3}$, orange and green trace, respectively) in
the region around the avoided level crossing. Clearly, the added charge density impacts the frequency tuning behavior. In
Fig.~\ref{fig:S7} (c), we plot the frequency difference between both branches $\Delta\omega$ for the three charging
configurations. Each curve yields a distinct minimum, which heralds the position of the avoided crossing. The minimal value of
$\Delta\omega$ corresponds to the splitting $\Omega_0$. Figure~\ref{fig:S7} (c) makes it immediately apparent that both the
position and the magnitude of the avoided crossing shift with increasing charge density. Figure~\ref{fig:S7} (d) depicts the
dependence of the level splitting $\Omega_0$ on the charge density. The splitting is found to increase almost linearly with the
charge density. The COMSOL simulation thus demonstrates a distinct effect of the charge density on the coupling strength between
the two mechanical modes.

This is verified in the experiment discussed in the main text (c.f. Fig.~\ref{fig:05}), where the RMS of the microwave field is
used to change the electrostatic environment sensed by the nanomechanical string resonator. This constitutes a proof-of-principle
demonstration of the sensing capabilities of the nanomechanical resonator. 

For a more quantitative analysis, we start from the theory precition that IAS is expected to resolve changes of order $10^{-4}
\Omega_0$, depending on the coherence time of the nanostring (see Fig.~3 of Ref.~\cite{Chowdhury2022}).  This can be conveniently
expressed in terms of a net charge density on the nanostring. Based on the slope of Fig.~\ref{fig:S7} (d), we expect a charge
sensitivity of approximately $26$\,Hz/(C/m$^3$).  The shift recorded in Fig.~\ref{fig:05} (b) thus corresponds to a charge density of
$180$\,C/m$^3$ or an equivalent of $1,400$ electrons added to the nanostring, comparable to other vibrating room-temperature
charge sensors~\cite{Lee2008}.  Using the estimated sensitivity of IAS~\cite{Chowdhury2022} of $10^{-4} \Omega_0$ the charge
sensitivity gives rise a minimum detectable charge of $5.65$\,C/m$^3$, corresponding to approximately $43$ electrons on the
nanostring. We envision that a series of straightforward improvements of the sample geometry, including a narrower nanostring
cross-section as well as a reduced electrode separation, will allow to detect a single electron being added to the nanostring. 

A key feature of this sensing scheme is its high time resolution. For the experiment described in the main text [$n=4$ Ramsey
fringes, $m=8$ iterations, see Fig.~\ref{fig:03} (c)], an estimate for $\Omega_0$ is obtained after approximately $1$ s. Reducing
the number of iterations (see Fig.~\ref{fig:05}), and taking advantage of an FPGA to apply pre-configured pulses, will allow to
reduce the measurement time to a few $100\,\mu$s. This is considerably faster than the comparably sensitive scheme for room
temperature charge sensing based on coupled nonlinear mechanical resonators~\cite{Wang2020} or a single diamond
spin~\cite{Dolde2011}. 

Another potential field of application of IAS vectorial scanning force sensing based on a vibrating nanowire~\cite{Gloppe2014,
Rossi2016a,Lepinay2017,Braakman2018,Braakman2019,Fogliano2021}. The vectorial scanning force sensor is a nanomechanical two-mode
sensor. It probes the effect of its electrostatic environment through a change of the splitting between the two normal modes in
the coupled reference frame of the nanowire, and thus relies on the same detection principle as IAS. The integration of IAS is
therefore straightforward. It promises to increase the sensitivity as a result of its robustness to noisy environments.

\end{appendix}

\end{document}